**Heat flux deposition pattern on the inner first wall of Tore Supra**

R. Mitteau, M. Chantant, P. Chappuis, D. Guilhem, M. Lipa

*Association Euratom-CEA, Département de Recherches sur la Fusion Contrôlée, CEA/Cadarache*
*F-13108  SAINT PAUL LEZ DURANCE CEDEX (France)*

**Abstract**

The Inner First Wall (IFW) is the main limiter in Tore Supra for long and powerful discharges. The original graphite IFW was partly damaged by 8 years of operation and a sector of 60° was replaced between 1995 and 1997 by CFC brazed elements. They are designed to accept twice as much flux (2 MW/m²) as the older ones and are expected to be more resilient. The new elements gave the opportunity to install diagnostics (thermocouples and langmuir probes) on a poloidal sector, to characterise the behaviour of the new IFW and especially the heat flux profile. The pattern doesn't follow the cosine law and is governed by strong perpendicular heat flux and abnormal deposition at the contact point between the limiter and the plasma. The most energetic shots of the last campaign showed that the IFW could be loaded with twice the power deposited during these shots.

## 1. INTRODUCTION

The Inner First Wall (IFW) covers 25 m² on the high field side of the vacuum chamber. The older elements (300° toroidal) have graphite tiles and the new one (60° toroidal) carbon fiber composite (CFC) tiles [1,2]. The total area of carbon based material wetted by the plasma is 11 m². It is composed of 1476 flat elementary sites bearing 6 tiles (fig. 1). The IFW has a polygonal shape fitting the toroidal vessel. It is actively cooled and reach thermal equilibrium in 30 seconds. The new 60° are positioned radially 2 mm ahead of the others. Thermocouples and langmuir probes were installed on a new 3°20 sector. These diagnostics will be described in the next section. Part of the experimental program of 1997 was devoted to the qualification of the new elements. For this purpose, long (25 s) and powerful (up to 6.1 MW) discharges where run, resulting in injected energies up to 170 MJ. The results on the IFW will be presented and interpreted.

## 2. INSTRUMENTATION

A 3°20 poloidal sector was prepared to bear specific instrumentation, thermocouples and langmuir probes (fig. 1). Thirteen thermocouples were bolted to the copper sheets acting as compliance layer between the CFC tile and the stainless steel heat sink. Due to the averaging effect of this high conductivity layer, the temperature measured by the thermocouple is representative of the average heat flux on the site. In addition to the thermocouples, 4 langmuir probes were installed between the sites to measure electron temperature and density. Other independent diagnostics are also used to monitor the IFW : an infrared camera (IR) and the calorimetric measurements on the cooling loop [3]. The resolution of the camera is better than one site (80 mm * 130 mm), but individual tiles aren't seen. The calorimetry is recorded by module (60° toroidal sectors of the machine) allowing to have a toroidal distribution of the incident power. These diagnostics are redundant which allows to cross-check the data and give good confidence in the results. One should however consider that IR and calorimetric measurements yield their best results for the most energetic shots, because short (<5 s) or less powerful (<2 MW) discharges don't produce enough heating of the plasma facing components.

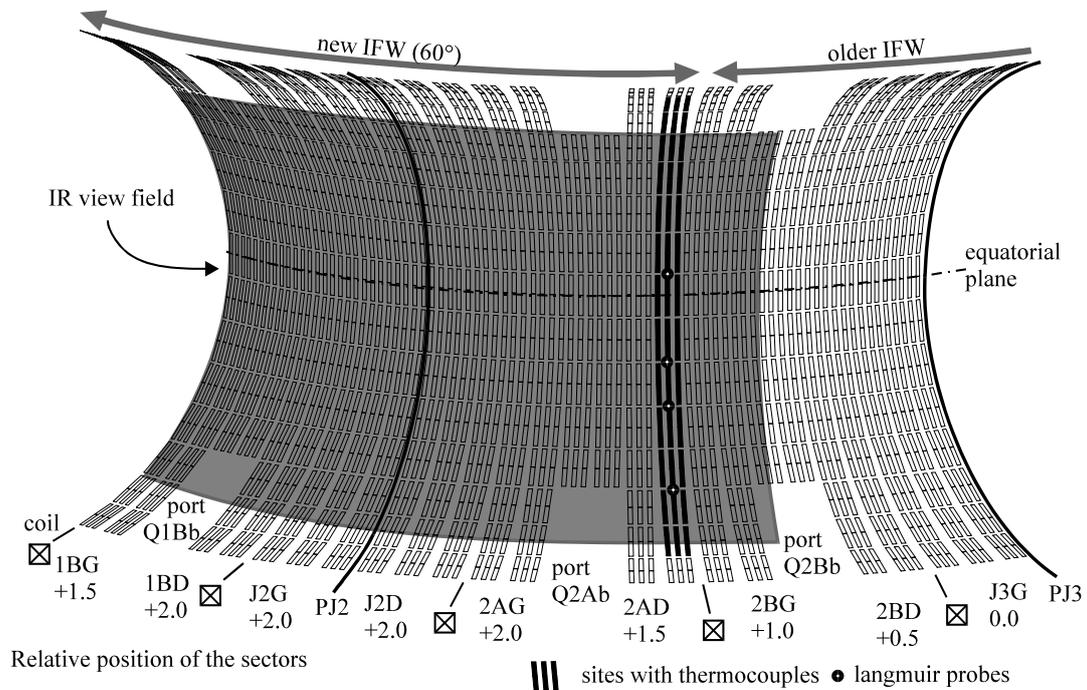

figure 1 : diagnostics of the Inner First Wall

## 3. OBSERVATIONS

Individual sites are seen on the infrared image of the IFW in operation (fig. 2). The heat pattern is regular. No hot spots caused by faulty brazed joints or cracked tiles are to be seen, except on the very right where a row of the older IFW overheats. The heat deposition is peaked on the equatorial plane between the toroidal coils. The ripple of the toroidal magnetic field governs this pattern. Although of minor magnitude on the inner side of the machine than on the outer side, the ripple still influence strongly the heat deposition on the IFW.

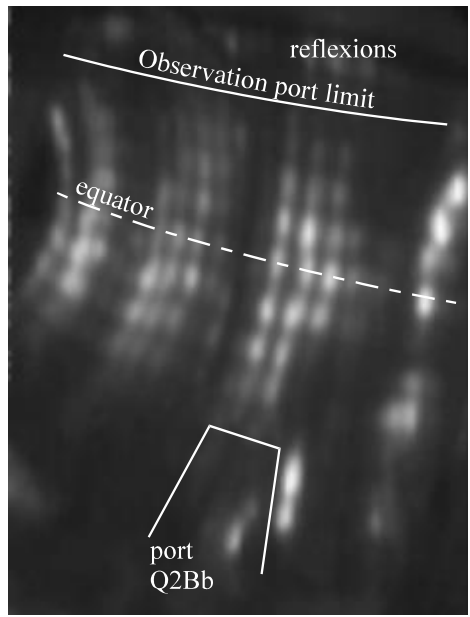

figure 2 : Infrared image of the IFW (shot #23871 at 25s), black is 150°C and white 650°C

The temperature of the thermocouples is recorded during the shots. The maximum increase of the temperature is shown with a continuous line fig. 3 as a function of the poloidal position for 3 shots (#23869, #23870 and #23871). The data are very much similar for the three shots. The water temperature of the cooling loop is 150°C. A general increase of the temperature is observed, making a kind of pedestal of around 35°C. This is observed on the thermocouples as well as on the infrared data which are plotted for #23871. A strong peak is observed on the equatorial plane, mainly on the middle site but also on his two immediate neighbours up and down. A dissymmetry can be noticed on both diagnostics, the lower part (electronic side) being somewhat more heated.

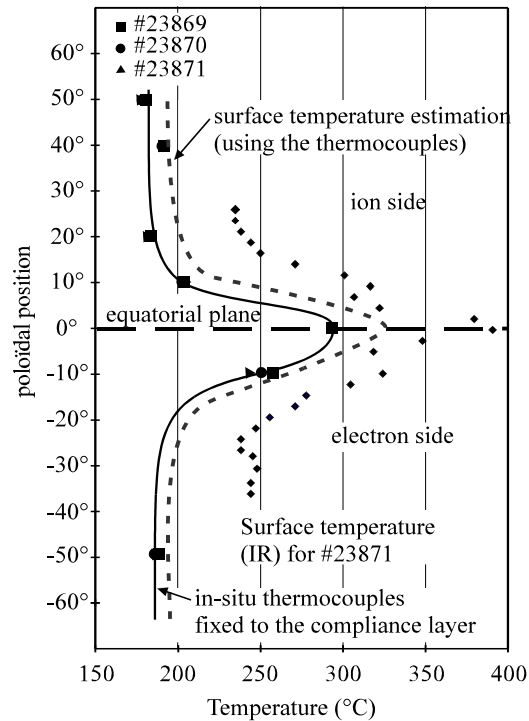

**figure 3 : Poloidal profile of the temperature**

The calorimetric measurements show that 2.2 MW are deposited on the IFW. The new sectors scraping one millimeter more than the other receive 22 % more than the average power per module.

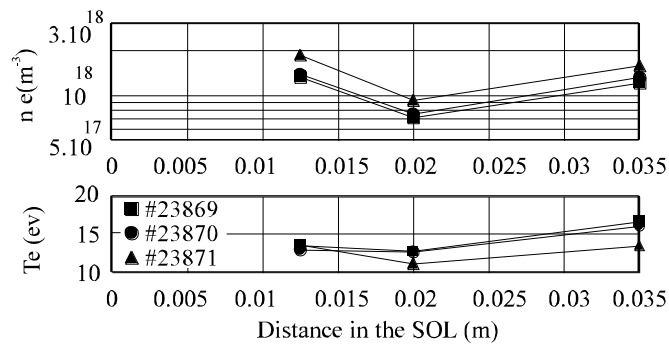

**figure 4 : fixed langmuir probes (located on the IFW at poloidal angles 5°, 15°, 25°, and 45° from the equatorial plane)**

Langmuir probes show a very flat distribution for a distance in the SOL larger than 10 millimeters (fig. 4). No measures are available in the equatorial plane due to a faulty probe so that the equatorial peak couldn't be analyzed in more details with electronic temperature and density.

## 4. DISCUSSION

From the copper layer and surface temperatures, an average heat flux on the site is evaluated using standard finite element calculations. The heat flux is also evaluated from temperature/time plots with the semi-infinite model during the first 5 seconds of the heating. All methods give comparable results and show that for the 2.2 MW extracted by the IFW, the average heat flux on the most heated site is 1.2 MW/m². This gives the confidence that the power on the IFW could be almost doubled for such shots without endangering it. A surface temperature is also evaluated from the average heat flux deduced from the thermocouples (fig. 3, dashed line). The discrepancy to the actual surface temperature measured with the infrared camera is explained by the fact that a fraction of the heat flux reaches the tiles with a grazing angle, thus heating mainly the leading edges and increasing the overall luminance seen by the camera. Regarding the continuous background, the radiative heat flux measured with bolometers and with the calorimetric measurement on the water cooled back plates can only account for one third of the temperature increase. The rest have to be associated to local plasma physics (perpendicular heat flux) which was already mentioned in an earlier study [4].

The equatorial peaking is also problematic. A classical $\lambda q$ can be deduced from the power ratio of the first element to the total power.

$$\lambda q = -\delta_1 \cdot \left( \ln(1 - \frac{P_1}{P}) \right)^{-1}$$

In that formula, $\lambda q$ is the e-folding length for the power deposition, $P_1$ the convective power scraped by the first element, P the total convective power of the SOL and $\delta_1$ the distance of the SOL intercepted by the first element (1 millimeter for the IFW. Two calculations are presented table 1, one in which the continuous background is maintained ("global $\lambda q$"), one in which it is subtracted from the profile ("contribution of the peak only"). In addition, the same principle is used on the toroidal profile with the calorimetry on the advanced new panels which are 1 millimeter beyond the older. Very narrow $\lambda q$ are found, which are not consistent with the usual one of 10 millimeter [5]. Similarly, the profile of the parallel convective heat flux can be calculated with convective code TOKAFLU [6]. Very short e-folding length have to be taken to make the profiles fit. Therefore, the peak cannot be explained by parallel convective heat flux.

**table 1 : lq evaluation**

| $\lambda q$ in millimeters | thermocouples | calorimetry |
|---|---|---|
| with continuous background "global $\lambda q$" | 3.8 | 4.4 |
| continuous background subtracted "peak only" | 1.2 | 1.9 |

## 5. CONCLUSIONS

The new CFC brazed panels of the IFW are a technological success. They extend the operating range of the IFW, which is today only limited by the older panels. The new sector allow a fair observation of the heat deposition pattern which can be described as a broad base "continuous background" caused by radiation losses and perpendicular heat flux and strong peaks located at the contact points between the plasma and the wall. All the diagnostics installed on the IFW confirm these observations. Modelisation of the heat pattern is possible with various codes (THOR for perpendicular heat flux, TOKAFLU for the equatorial peak). Although the quantitative physical justification lacks, the experimental pattern could be reproduced by adjusting the relative contributions of the different models with adequate diffusion coefficient in the perpendicular direction and $\lambda q$.

**REFERENCES**

[1] Lipa, M., et Al. "Development and fabrication of a new generation of CFC-brazed plasma facing component for Tore Supra", report EUR/CEA-FC-1550, July 1995.
[2] Schlosser, J. at Al. , "In service experience feedback of the Tore Supra actively cooled inner first wall", Fusion Engineering and Design **27**(1995)203-209.
[3] Surle, F., et Al. "Calorimetric measurements of energy deposition in Tore Supra", presented at the 19th SOFT, Lisbon, 16-20 September 1996.
[4] Seigneur, A. et Al., "The use of large surface area for particle and power deposition", presented at the 20th EPS, Lisbon, 16-30 July 1993.
[5] Guilhem. D. et Al., "Actively cooled pump limiters and power scrape-off length measurements in Tore Supra", JNM **196-198**(1992)759-764.
[6] Mitteau, R. et Al, "Heat flux deposition on plasma facing components using a convective model with ripple and Safranov shift", presented at 13th PSI conference, 18-22 may 1998, San Diego.